\documentclass[journal=ancham,manuscript=article]{achemso}
\usepackage[version=3]{mhchem} % Formula subscripts using \ce{}
\usepackage{ulem}
\usepackage{graphicx}
\usepackage{epstopdf}
\usepackage{amsmath}
\usepackage{achemso}
\usepackage{color}
\usepackage{physics}
\usepackage{MnSymbol}

\newcommand{\meant}[1]{\langle{#1}\rangle}
\newcommand{\meanN}[1]{\lbrace{#1}\rbrace}

\newcommand{\etal}{\textit{et al.~}}
\newcommand{\ie}{\textit{ie.~}}

\newcommand{\etc}{\textit{etc}}
\newcommand{\reffig}[1]{Fig. \ref{#1}}

\author{Antoine Riaud}
\affiliation[ASIC]{ASIC and System State Key Laboratory, School of Microelectronics, Fudan University, Shanghai 200433, China}
\alsoaffiliation[Sorbonne]{INSERM UMR-S1147, CNRS SNC5014, Paris Descartes University, Equipe labellis\'{e}e Ligue Nationale contre le cancer, Paris, France}
\email{antoine_riaud@fudan.edu.cn}
\author{Anh L. P. Thai}
\altaffiliation{The Physics of Living Matter Group, Department of Physics and Materials Sciences, University of Luxembourg}
\affiliation[Sorbonne]{INSERM UMR-S1147, CNRS SNC5014, Paris Descartes University, Equipe labellis\'{e}e Ligue Nationale contre le cancer, Paris, France}
\author{Wei Wang}
\affiliation[ASIC]{ASIC and System State Key Laboratory, School of Microelectronics, Fudan University, Shanghai 200433, China}
\author{Valerie Taly}
\affiliation[Sorbonne]{INSERM UMR-S1147, CNRS SNC5014, Paris Descartes University, Equipe labellis\'{e}e Ligue Nationale contre le cancer, Paris, France}
\email{valerie.taly@parisdescartes.fr}

\title
  {Mechanical characterization of cells and microspheres sorted by acoustophoresis with in-line resistive pulse sensing} 

\abbreviations{IR,NMR,UV}

\begin{document}
%\linenumbers
%\begin{tocentry}

%\end{tocentry}

\begin{abstract}
Resistive Pulse Sensing  (RPS) is a key label-free technology to measure particles and single-cell size distribution. As a growing corpus of evidence supports that cancer cells exhibit distinct mechanical phenotypes from healthy cells, expanding the method from size to mechanical sensing could represent a pertinent and innovative tool for cancer research. In this paper, we infer the cells compressibility by using acoustic radiation pressure to deflect flowing cells in a microchannel, and use RPS to sense the subpopulations of cells and particles at each acoustic power level.  We develop and validate a linear model to analyze experimental data from a large number of particles. This high-precision linear model is complemented by a more robust (yet less detailed) statistical model to analyze datasets with fewer particles. Compared to current acoustic cell phenotyping apparatus based on video cameras, the proposed approach is not limited by the optical diffraction, frame rate, data storage or processing speed, and may ultimately constitute a step forward towards point-of-care acousto-electrical phenotyping and acoustic phenotyping of nanoscale objects such as exosomes and viruses.
\end{abstract}

\section{Introduction}

Resistive pulse sensing (RPS) is a key method for the label-free analysis of cells and nanoparticles \cite{guo2015demand}. It measures the size of particles by monitoring the electrical resistance of a channel filled with electrolyte such as Phosphate Buffer Saline (PBS). When an insulating particle travels through this channel (or pore), the resistance changes by an amount proportional to the particle volume. This method is faster than video analysis and is not limited by optical diffraction. This makes it pertinent to analyze viruses \cite{harms2011nanofluidic,zhou2018characterization} colloids \cite{weatherall2016pulse,willmott2018tunable}, macromolecules \cite{billinge2013monitoring} and especially DNA \cite{clarke2009continuous}. For all these advantages, RPS is a mainstream method with well-established standards adapted for medicine and diagnostics based on liquid biopsies\cite{braylan1978cell,chapman1981cell}.

In recent years, the accumulation of evidences showing that cancer cells exhibit distinct mechanical phenotypes from healthy cells \cite{suresh2007biomechanics,darling2015high} has prompted a major research effort to add mechanosensing capabilities to the RPS framework. The most common strategy \cite{koutsouris1988determination,zheng2012high,zheng2015decreased,zhou2017characterizing} is to use a pair of constrictions of different widths. The widest constriction is slightly larger than the cell diameter and measures the cell size while the smallest one is slightly narrower than the cell. The time needed for the cell to deform and travel through the second constriction may then be fed into biomechanical models to estimate the cell deformability \cite{raj2017combined,ye2018relationship}.  Two major requirements of this squeezing method are (i) that the constriction size must be very close to the cell size, which requires specific devices for each different cell size and make it unsuitable for complex mixtures of different cells (such as whole blood and heterogeneous populations\cite{ruban2015mononuclears}), and (ii) that the objects traveling through the channel must all be highly deformable, which excludes the study of solid microparticles and most nanoparticles (including viruses and exosomes). Indeed, the shear due to cytoplasm flow in deforming cells is proportional to $1/a^2$\cite{ye2018relationship}, with $a$ the cell radius, and by analogy it is inferred that viral and exosome components could not deform enough to cross a narrow constriction and provide meaningful elasticity data.

While deformation is a quasi-static measurement of cell mechanical properties, acoustic characteristics such as sound speed and compressibility complement this picture with a high-frequency viewpoint. Such measurement is commonly achieved by acoustophoresis, \ie by tracking the migration of objects due to acoustic forces \cite{hartono2011chip}. Compared to deformation measurements, acoustophoresis characterization is applicable to both solid and soft particles and is contactless, which minimizes cross-contamination risks. A major shortcoming of this characterization is that the migration speed depends strongly on the particle diameter, which has to be evaluated externally, for instance using RPS\cite{hartono2011chip}. A promising alternative is the isoacoustic method\cite{augustsson2016iso} but it requires an elaborated optical setup and modified media. Both methods also rely on a microscopy setting for video analysis which restricts the measurement throughput, is limited by optical diffraction and precludes point-of-care applications.  

In this paper, we combine the well-accepted RPS method for the measurement of particle size to acoustophoresis for the measurement of particle compressibility. We first introduce a theoretical model that relates the particle deviation to the acoustic field intensity and the particle size, density and compressibility. Hence, depending on the acoustic power level, different populations of particles can be sorted. The size of each of these populations is then measured by an in-line RPS chip. Eventually, the signal is numerically analyzed to recover the particle compressibility depending on its density.

After introducing the necessary theoretical background for each component of the experiment (acoustophoresis chip, resistive pulse sensing chip and signal analysis) and detailing the experimental protocol, we coupled an acoustophoresis and an RPS chips for a proof-of-concept experiment involving polystyrene microspheres and Jurkat cells (a common model for blood cancer already studied with the RPS technology alone). Our experiments not only yield to the compressibility and the size distribution of the polystyrene microspheres and Jurkat cells but also reveal that the polystyrene microspheres formed doublets that could be distinguished from Jurkat cells based on their compressibility.

\section{System principle and theory}

\begin{figure}
\includegraphics[width=80mm]{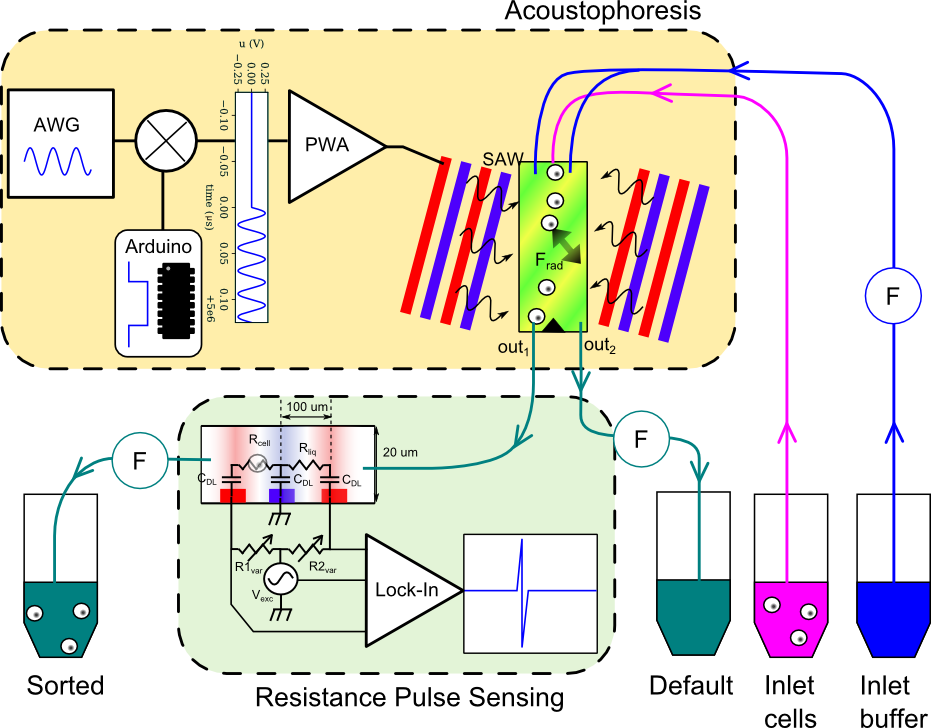}
\caption{Experimental setup. The experiment combines an acoustostophoresis chip (top-center) and a RPS chip (bottom-center). On the bottom-right, two pressurized vials containing a buffer fluid (PBS) and the sample to analyze (cells and/or microspheres) supply the acoustophoresis section of the experiment (top). This section comprises a microfluidic chip (top-right), a pair of interdigitated transducers and a power-modulated ultrasonic power supply (top-left). Depending on the acoustic power, the particles are directed to a default outlet (bottom-right) or to the resistance pulse sensing chip (bottom-center). The flow rates in the system are controlled by three precision flowmeters.}
\label{fig: general_schematic}
\end{figure}

The proposed system combines acoustophoresis to measure particle compressibility and resistive pulse sensing to obtain particle size (\reffig{fig: general_schematic}). By modulating the acoustic power level, the acoustophoresis chip sorts various populations of particles. The relation between the acoustic power and the deviation of particles is established at the beginning of this section. It is shown that the particle compressibility can only be computed if the size of the particles is known. To obtain the particle size, the sorted particles are guided towards another chip equipped with an RPS sensor. In principle, knowing the acoustic power and the particle size should yield the acoustic contrast of the particle and thus its compressibility in a straightforward fashion. However, the two chips were designed as stand-alone and are connected by a short yet non-negligible capillary tubing that generates a lag between the sorting and detection events. The signal processing when accounting for this delay is non-trivial and is discussed at the end of this section.

\subsection{Tilted-angle standing SAW acoustophoresis}

\begin{figure}
\includegraphics[width=80mm]{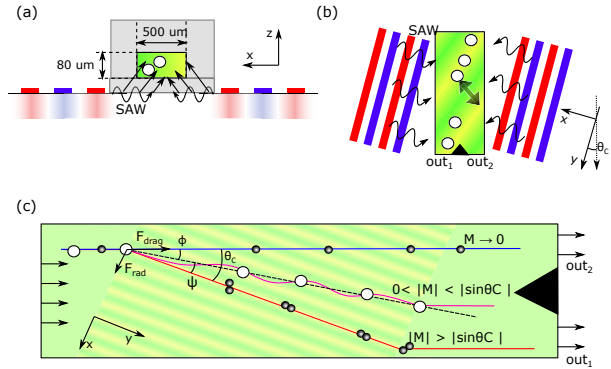}
\caption{Sorting chip schematic. \textbf{(a)} cross-section: the IDTs are located on each side of the disposable PDMS chip. The generated SAWs (wavy arrows) travel freely along the solid surface and radiate once they reach the PDMS. The $x$ direction in the drawing corresponds to the crystallographic $\mathbf{X}$ direction. The generated bulk waves are then transmitted into the liquid (straight arrows). \textbf{(b)} top view: the particles carried by the fluid are deviated by the acoustic radiation force. The blue and red colors indicate the (oscillating) electric potential while the green to yellow color gradient refers to the (oscillating) acoustic pressure field. \textbf{(c)} Three scenarios of particle deviation depending acoustic radiation to drag force ratio $M$ (Eq. \eqref{eq: M}): when $M\rightarrow0$, the particles follow the flow ($\psi=\theta_C$), whereas when $|M|>|\sin\theta_C|$ the particles are locked along the acoustic wavefronts ($\psi=0$). At intermediate values, the particles follow a striated path forming an angle $\psi$ with the wavefronts given by Eq. \eqref{eq: Pelton}.}
\label{fig: schem acoustophoresis}
\end{figure}

At high power, acoustic waves generate a steady stress called acoustic radiation pressure. The resulting force is used in acoustophoresis experiments to displace particles \cite{riaud2017selective,baudoin2019folding}. Since the migration speed depends on the particles size, density and compressibility\cite{williams2017acoustophoretic}, acoustophoresis is routinely used for sorting small objects in microfluidic channels\cite{petersson2007free,augustsson2012microfluidic,muller2013ultrasound,grenvall2014two,augustsson2016iso}.

A widespread technology to generate the acoustic field is using surface acoustic waves (SAW) as shown in \reffig{fig: schem acoustophoresis}(a). At the center of the picture, cells suspended in their culture medium flow through a PolyDiMethylSiloxane (PDMS) channel. This channel is placed between a pair of interdigitated transducers (in blue and red) that generate two counter-propagative surface acoustic waves (oscillating arrows). These waves propagate along the solid surface until they reach the microchannel base. At this stage, the SAW interfere to form a standing surface acoustic wave and the vertical component of the surface wave vibration radiates into the PDMS and then in the channel as a bulk acoustic wave (BAW). A considerable advantage of using SAW is the possibility to use cheap disposable microchannels that can be detached from the ultrasonics transducer in order to minimize cross-contamination risks\cite{kishor2015characterization,guo2015reusable,Ma2016}.

Under the action of the acoustic wave, the cells are attracted towards the pressure nodes. At best, this allows a separation distance of a quarter-wavelength. In order to overcome this limitation, Collins \etal  \cite{collins2014particle} proposed using tilted-angle SAW to deflect the particles as shown in \reffig{fig: schem acoustophoresis}(b). In this tilted configuration, the trapped particles will travel along the acoustic nodes while the drifting particles will follow the flow more closely (cf. analytical model thereafter). This tilted configuration has later been improved by Ding \etal \cite{ding2014cell} to sort circulating tumor cells from peripheral blood mononuclear cells (PBMC) \cite{li2015acoustic} and exosomes from whole blood \cite{wu2017isolation} and is adopted here.

The acoustic force $\mathbf{F}_{rad}$ due to a standing SAW reads:
\begin{eqnarray}
\mathbf{F}_{rad} &=&  - \frac{4\pi a^3}{3} \mathbf{k}_X \mathcal{E} \Phi \sin(2 \mathbf{k}_X\cdot \mathbf{r}), \label{eq: Force}\\
\Phi &=& f_1 - \frac{3}{2}f_2\cos(2\theta_R),   \label{eq: SAW contrast factor}\\
\mathcal{E} &=& \frac{1}{2}\kappa_0 {p_{RMS}}^2 \label{eq: E_density}\\
p_{RMS} &=& \frac{\rho_0\omega^2 u_{RMS}}{k_L\cos\theta_R} \label{eq: pRMS}
\end{eqnarray}
with $a$ the particle radius, $\mathbf{k}_X$ the SAW wave-vector, $k_L$ the BAW wavenumber (in the liquid), $\omega$ the SAW angular frequency, $\mathcal{E}$ the acoustic energy density, $p_{RMS}$ the pressure fluctuation root-mean-square (RMS) of the BAW, $u_{RMS}$ the RMS of the SAW vertical oscillations, $\Phi$ the acoustic contrast between the particle and the fluid, and $\mathbf{r}$ the position vector of the particle. The acoustic contrast factor, given by Eq. \eqref{eq: SAW contrast factor}, depends on the propagation angle of the radiated SAW $\theta_R\simeq 22^o$ (Rayleigh angle) and the monopolar $f_1 =  1- \frac{\kappa_p}{\kappa_0}$ and dipolar $f_2 =  \frac{2(\rho_p - \rho_0)}{2\rho_p+\rho_0}$ scattering coefficients of the particle. $\kappa_p$ and $\kappa_0$ stand for the particle and fluid compressibility respectively, and $\rho_p$ and $\rho_0$ the particle and fluid density respectively. As pointed out by Simon \etal\cite{simon2017particle}, the SAW acoustic contrast differs from its BAW value by a factor $-\cos(2\theta_R)\simeq -0.67$, that is approximately the opposite of the unitary value needed to recover the usual contrast factor $\Phi = f_1 + \frac{3}{2}f_2$ obtained when  $\theta_R\rightarrow\frac{\pi}{2}$. 

The acoustic radiation force is balanced by the drag force:
\begin{equation}
\mathbf{F}_{rad}+\mathbf{F_{drag}}=\mathbf{0}. \label{eq: force_balance}
\end{equation}
Neglecting particle acceleration, and for particles far away from channel walls, the drag force reads:
\begin{equation}
\mathbf{F_{drag}} = 6\pi\eta a(\mathbf{v_F}-\mathbf{v_p}), \label{eq: drag}
\end{equation}
where $\mathbf{v_F}$ and $\mathbf{v_p}$ stand for the flow and particle speed respectively, and $\eta$ is the dynamic viscosity. 

In previous studies of tilted-angle standing SAW acoustophoresis, no solutions to Eq. \eqref{eq: force_balance} were available and it had to be integrated numerically. However, the analog optical problem was previously solved by Pelton \etal\cite{pelton2004transport} by a clever change of coordinates. They showed that the particle travels with an angle $\psi$ relatively to the wavefronts (see \reffig{fig: schem acoustophoresis}(c)):
\begin{eqnarray}
\tan\psi &=& \left\lbrace
\begin{array}{lr}
0, & |\sin\theta_C| < |M| \\
\frac{\sqrt{\sin^2\theta_C-M^2}}{\cos\theta_C}, & |\sin\theta_C|>|M|
\end{array}
\right.\label{eq: Pelton}\\
M &=& \frac{2a^2\Phi k_X\mathcal{E}}{9\eta v_F}. \label{eq: M}
\end{eqnarray}
Interestingly, Pelton \etal also derived a relatively simple analog of Eq. \eqref{eq: Pelton} for nanoparticles that accounts for Brownian diffusion.

\begin{figure}
\includegraphics{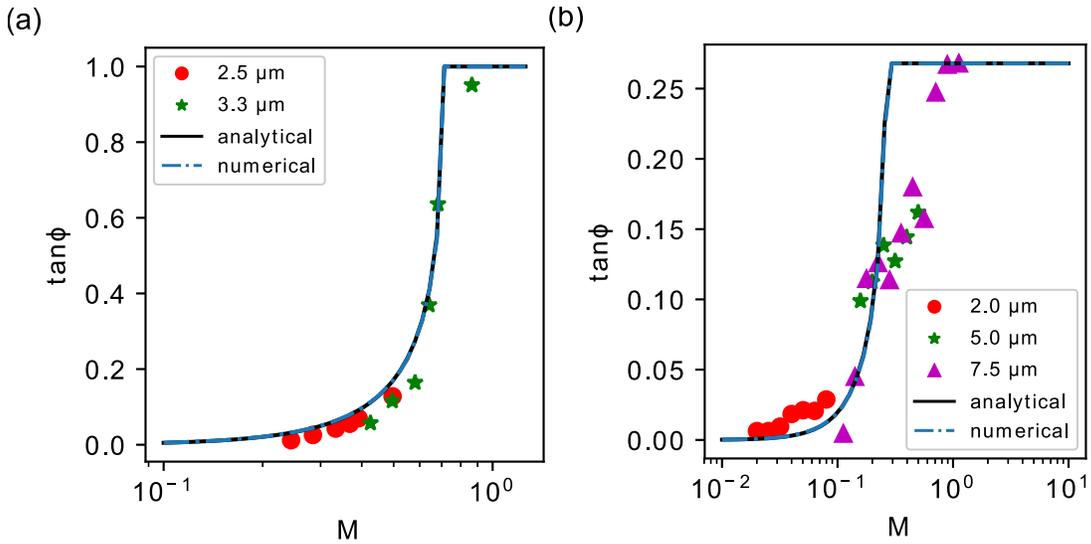}
\caption{Comparison of previously published deviation angle of polystyrene microspheres of various radii with the numerical integration of equation \eqref{eq: force_balance} and its analytical solution given by Eq. \eqref{eq: Pelton}. \textbf{a}: Adapted from Collins \etal  (Fig. 7B)\cite{collins2014particle}, \textbf{b}: Adapted from Ding \etal (Fig. S1) \cite{ding2014cell}. In both cases, the SAW magnitude was assumed proportional to the actuation voltage and this conversion coefficient was the only fitting parameter (one conversion coefficient was regressed for each figure). The angle $\phi = \theta_C-\psi$ was deemed closer to experimental concerns and thus more convenient for comparisons.}
\label{fig: comparison_pelton}
\end{figure}

In the deterministic (non-Bronwnian) case, this deviation angle only depends on the tilt angle $\theta_C$ and the dimensionless migration speed $M$ given by Eq.\eqref{eq: M}. When $|\sin\theta_C| < |M|$, the particles follow the acoustic wavefronts (locked mode) whereas for smaller values of $|M|$ the particles travel more tangentially to the flow (\ie $\tan\psi \rightarrow \tan\theta_C$). Remarkably, $\tan\psi$ is independent of the sign of $M$ so that particles with positive and negative acoustic contrast follow the same trajectory. Since these theoretical results were previously unknown to the field of acoustics, we first confirmed them against previously published data in \reffig{fig: comparison_pelton}. The perfect match between analytical and numerical results validates the calculations, while the good agreement with experimental results supports the validity of this opto-acoustic analogy.

According to Eq. \eqref{eq: Pelton}, the particles trajectory depends only on the parameter $M$. Since $M$ depends on the particle radius and acoustic contrast, previous studies\cite{hartono2011chip} were unable to obtain the compressibility directly and had to assume a given particle radius or use an average value obtained by an independent measurement instead. Hence, one experiment had to be performed for each particle size, and the heterogeneity in size was difficult to take into account. Here, the particle size is directly measured after sorting using the RPS chip. 

\subsection{Resistive pulse sensing}

\begin{figure}
\includegraphics[width=60mm]{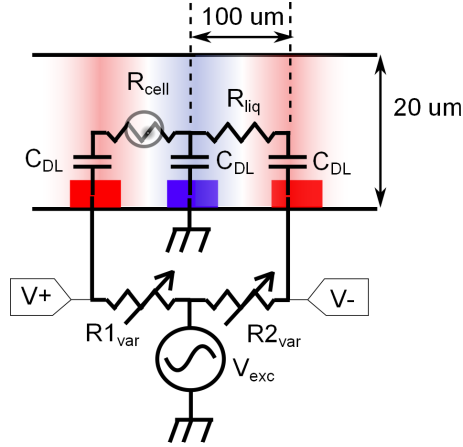}
\caption{Electrical schematic of the sorting chip. The cells and microspheres flow through the 200 $\mu m$ long constriction where they are sensed by the electrode triplet. This channel forms half of Wheatstone bridge completed by a variable resistor. The 50 kHz AC excitation allows to bypass the capacitive nature of the electrical double layer.}
\label{fig: schem RPS}
\end{figure}

Resistive pulse sensing works by monitoring the electrical resistance of a channel containing a conductive solution (\reffig{fig: schem RPS}). In the absence of particles, a voltage applied between the blue and red electrodes generates a baseline electrical current. When an insulating particle travels through this channel, it blocks some of the electrical current (\ie the channel electrical resistance $R_{CH}$ increases). The method is also able to determine biological cell radii thanks to the cells membrane that blocks the electrical current. For a channel of hydraulic diameter $D_H$ and length $L$\cite{coulter1953means}, the resistance increase $\Delta R_{CH} = R_{liq}-R_{cell}$ reads:

%\begin{multline}
%\left|\frac{\Delta R_{CH}}{R_{CH}}\right| = \frac{8 a^3}{L{D_H}^2}\left[\frac{{D_H}^2}{2L^2}+\frac{1}{\sqrt{1+\left(\frac{D_H}{L}\right)^2}}\right]\lbeq
%\times F(8a^3/{D_H}^3),\label{eq: res. coulter}
%\end{multline}
\begin{equation}
\left|\frac{\Delta R_{CH}}{R_{CH}}\right| = \frac{8 a^3}{L{D_H}^2}\left[\frac{{D_H}^2}{2L^2}+\frac{1}{\sqrt{1+\left(\frac{D_H}{L}\right)^2}}\right] F(8a^3/{D_H}^3),\label{eq: res. coulter}
\end{equation}
where $F$ is a correction factor close to unity when $a < 0.25D_H$\cite{deblois1970counting} and the hydraulic diameter is given by $D_H = \sqrt{\frac{4wh}{\pi}}$ with $w$ the channel width and $h$ its height. 

A major problem when working with cells and smaller particles is that miniaturized electrodes become a current bottleneck due to the insulating electrical double layer (depicted as capacitors $C_{DL}$ in \reffig{fig: schem RPS}). Hence, measurements of continuous current are flawed due to this parasitic capacitance. This limitation is mitigated by working with an alternative current. Another advantage of alternating current is to protect the electrodes from electrolysis.

In order to improve the signal-to-noise ratio, the channel electrical resistance is monitored by comparing two symmetric electrodes (shown in red), which form an electrical half-bridge. Measurement errors are further minimized by completing a 4-resistors Wheatstone bridge to obtain the differential signal directly:

\begin{equation}
V_+-V_- \simeq \frac{\Delta R_{CH}}{R_{CH}+R_1}V_{exc},\label{eq: Wheatstone}
\end{equation}

This signal is subsequently amplified by lock-in amplifier and digitized for numerical processing.

\subsection{Signal analysis}

Depending on the acoustic power level, different populations of particles are deflected towards the RPS chip where the particle size is evaluated. In principle, knowing the dimensionless migration speed $M$ and the particle size should yield the acoustic contrast of the particle and thus its compressibility. In practice, the two chips were designed as stand-alone and are connected by a short yet non-negligible capillary tubing. Due to this additional distance, there is a lag of several minutes between the application of the acoustic power and the detection of particles in the RPS chip. This situation is further complicated by the Taylor dispersion so that particle arrival time itself spreads over tens of seconds. Furthermore, due to experimental constraints detailed thereafter, the acoustic power levels had to be changed randomly over relatively short timescales. In order to process the time-delayed RPS signal and obtain the particles compressibility, we propose two approaches. The first strategy is closely related to a physical model of the particle transport in the capillary model (linear model) while the second one is purely statistical and focuses on the acoustic power levels that preceded the detection of a particle. 

%In both analyses, we consider a population of $N$ particles/cells and for each particle $i$ we record the time-dependent power levels $\mathcal{E}_i(\tau) = \mathcal{E}(t_i-\tau)$ that preceded the arrival of the particle/cell at time $t_i$ by a delay $\tau$. Each positive population is completed by a symmetric control population of negative particles, which are $N$ measurement time-points $t_i$ sampled randomly when no particles/cells were observed. Similarly, we record the associated time-dependent power levels $\mathcal{E}_i(\tau)$. Next, the time-dependent power series is discretized in quantified power levels $\mathcal{E}_k\in\lbrace\mathcal{E}_0..\mathcal{E}_p\rbrace$ and discrete delays $\tau_j$.

\subsubsection{Linear method}

In the physical approach, we consider the probability $g$ to observe the particle $i$ exiting the tubing at a specific time $t_i$. In physical terms, $g$ is the impulse response of the tubing for particle transport. A fundamental assumption of our model is that particles do not interact with each other, such that the arrival time distribution of the particles only depends on the particle (its size) and whether it was sorted or not (acoustic contrast and acoustic energy density). Another assumption is that the acoustic contrast distribution of each kind $K$ of particle do not overlap. Here, kind is purposefully vague as it may refer to the material (for polystyrene microspheres), or to the cell type, strain, ~\etc.

The experiments were conducted with a finite set of $p+1$ acoustic power levels ($\mathcal{E}_k\in\lbrace\mathcal{E}_0..\mathcal{E}_p\rbrace$). The time was also discretized into $n_t$ periods so that the delay $\tau$ between sorting and observation belongs to $\{\tau_1...\tau_{n_t}\}$. Assuming that the energy density was always zero except between $t_i - \tau_j$ and $t_i - \tau_{j+1}$ where it reached $\mathcal{E}_k$, we define $g_{jk}^{(K)}$ as the probability of observing a particle of kind $K$. Thus, the total probability of observing a particle of kind $K$ at time $t_i$ knowing all the sequence of power levels reads: 
\begin{eqnarray}
\hat{B_i}^{(K)} &=&  \sum\limits_{j = 0}^{n_t}\sum\limits_{k = 0}^{p} g_{jk}^{(K)}\delta_{ij}^k, \label{eq: total proba type}\\
\delta_{ij}^k &=& \left\lbrace
\begin{array}{lc}
1 & \mathcal{E}(t_i-\tau_j) = \mathcal{E}_k, \\
0 & \mathtt{otherwise}.
\end{array}\right. \label{eq: delta}
\end{eqnarray}
In Eq. \eqref{eq: delta}, $\delta_{ij}^k$ encodes reconstruction of the history of the acoustic energy density that preceded the detection of the particle $i$.

Since the power levels were randomly sampled from a uniform distribution, the total probability to observe a particle at time $t_i$ knowing all the sequence of power levels reads:
\begin{eqnarray}
\hat{B_i} &=&  \sum\limits_{j = 0}^{n_t}\sum\limits_{k = 0}^{p} g_{jk}\delta_{ij}^k, \label{eq: total proba}\\
g_{jk} &=& \sum\limits_{K} x_K  g_{jk}^{(K)},
\end{eqnarray}
where $x_K$ the fraction of particles of kind $K$. Compared to \eqref{eq: total proba type}, this equation indicates that the function $g_{jk}$ should exhibit one peak for each kind of particle even if they share the same acoustic contrast. Introducing the linear index $\alpha = j+n_tk$, the unknown $g$-function is obtained by comparing the model predictions $\hat{B}=\mathcal{D}G$ to the ground truth $B$ for $N$ particle detection events and $N$ controls (randomly sampled signals when no particles were detected). This comparison is done by minimizing the Euclidian distance $\|\hat{B}-B\|^2$ with the constraint $g_\alpha\geq0$:

\begin{eqnarray}
\mathcal{D} &=&  
\left[\begin{array}{ccc}
\delta_1^1&\cdots&\delta_{n_t (p+1)}^1\\
\vdots& & \vdots\\
\delta_1^{2N}&\cdots&\delta_{n_t (p+1)}^{2N}
\end{array}
\right], \label{eq: linear system D}\\
G &=&  
\left[\begin{array}{c}
g_1\\
\vdots\\
g_{n_t (p+1)}
\end{array}\right],
\quad
B =  
\left[\begin{array}{c}
B_1\\
\vdots\\
B_{2N}
\end{array}\right],
\label{eq: linear system G}\\
B_i &=& \left\lbrace
\begin{array}{lc}
1 & \mathtt{particle}, \\
0 & \mathtt{control}.
\end{array}\right.
\label{eq: linear system B}
\end{eqnarray}
which allows determining the impulse response for each type of particle at each power level. This detailed information can then be used to determine the minimum acoustic power to deflect them, and thus the acoustic contrast and particle compressibility.

\subsubsection{Statistical method}

When the number of particles is too low, the linear model is badly conditioned and yields inaccurate results. If the sorting threshold is the only valuable measurement to extract from the RPS dataset, the statistical method detailed thereafter tends to fare better. Compared to the previous model, an additional assumption is that there exists a sharp threshold $\mathcal{E}_{\min}$ below which no particles are deviated and above which all particles are sorted. Assuming that a discrete time-series $\mathcal{E}(t_i-\tau_j)$, $j\in\{1..n_t\}$ of random power levels yielded a particle $i$, we know that at least one of those power levels exceeded $\mathcal{E}_{\mathtt{min}}$. Reversely, if no particle was sorted we may assume that either no particle was present in the sorting section or that the threshold power has not exceeded at the critical time when the particle was in the sorting section. The former hypothesis becomes overwhelmingly more likely if the number of particles is small, which allows lifting the ambiguity. The problem then becomes analog to rolling an $s$-sided dice $n_t-1$ times and then rolling once a different dice that only has markings above a threshold $\mathcal{E}_{\mathtt{min}}$. The average result of the rolls will be higher than if the unbiased dice had been kept for all the rolls. Similarly the mean of the power series preceding particles detection should deviate from the mean of the power applied to the microchannel:
\begin{equation}
\meanN{\meant{\mathcal{E}_i}} = \frac{(n_t-1)\mu(\mathcal{E}) + \mu(\mathcal{E}\geq\mathcal{E}_{min})}{n_t},
\label{eq: stat}
\end{equation}
where $\meanN{x} = \frac{1}{N}\sum\limits_{i=1}^{N} x_i$ denotes the mean of quantity $x$ over all the particles, $\meant{x} = \frac{1}{\tau_{n_t}-\tau_1}\sum\limits_{j=1}^{n_t} x(\tau_j)$ is the time-average of the quantity $x$, $\mu(\mathcal{E})$ denotes the expected value of the s-sided dice and $\mu(\mathcal{E}>\mathcal{E}_{\mathtt{min}})$ the expected value of the biased dice:
\begin{eqnarray}
\mu(\mathcal{E}) &=&  \frac{1}{p+1}\sum\limits_{k=0}^{p}\mathcal{E}_k, \label{eq: dice_means}\\
\mu(\mathcal{E}\geq\mathcal{E}_{min}) &=&  \frac{1}{1+p-p_\mathtt{min}}\sum\limits_{k=p_\mathtt{min}}^{p}\mathcal{E}_k, \label{eq: dice_means_min}
\end{eqnarray}
In our experiments, the acoustic power levels are regularly spaced ($\mathcal{E}_k = k\mathcal{E}_0$). Combining Eq. \eqref{eq: stat} to \eqref{eq: dice_means_min} and after some algebraic manipulation, we get the shift between the average acoustic power levels that preceded the detection of a particle and those that do not:
\begin{equation}
\meanN{\meant{\mathcal{E}_i}} - \mu(\mathcal{E}) = \frac{\mathcal{E}_{min}}{2 n_t}, \label{eq: stat_shift}
\end{equation}
Eq. \eqref{eq: stat_shift} clearly illustrates the trade-offs of this statistical method: the acoustic power history must be long enough to capture the power level that triggered the sorting and subsequent detection of the particle, but it should not be too long as this tends to dilute the information.

\section{Materials and methods}

\subsection{Materials}

An aqueous suspension of 7.32 $\mu$m diameter PS microspheres (FS06F/9559) was purchased from Bangs Laboratories. According to the manufacturer, the PS beads density is 1062 $kg/m^3$.

Jurkat cells were prepared as described by Fernandez \etal \cite{fernandez20176}. The diameter of Jurkat cells is approximately $11.5\pm1.5\mu$m and their density was assumed similar to lymphoblasts\cite{zipursky1976leukocyte} (1075 $kg/m^3$).

Cells and PS microspheres suspensions were mixed together to a final number density of 0.5 million microspheres and 0.5 million cells/mL. This number density was chosen so that at most one cell or one particle was in the RPS sensor at any given time.

\subsection{Acoustophoresis chip and transducers}

The SAW transducer was a two-side polished 3'' diameter Y-128$^o$ cut of LiNbO$_3$ crystal equipped with a pair of interdigitated transducers. The electrode width and gap were set to 25 $\mu m$ in order to become resonant at the 40 MHz excitation frequency. The transducers were positioned to generate an X-propagating SAW (velocity $c_{SAW}=3990$ m/s). 
Disposable microfluidic chips made of PDMS were prepared by soft-lithography. The chips bottom were closed by a 100 $\mu$m thick membrane. According to profilometer measurements, the sorting section was 4 mm long, 500 $\mu$m wide and 80 $\mu$m high. It made a $3.4^o$ angle with the IDT. The fabrication process is detailed in the supplementary information.

The cells and microspheres were introduced at the channel center at a flow rate of 4 $\mu$L/min, while a buffer flow with a flow rate of 9 $\mu L$/min was added symmetrically to focus the particles before sorting. Such flow rate was chosen as an acceptable compromise between slower flow rates that yield an easier sorting and high flow rates less prone to sedimentation issues. The flow was supplied by a pressure-based microfluidic flow controller (MFCS-EZ, Fluigent) and the flow rates were monitored with three microfluidic flow sensors (FRP, Fluigent). The pressure was then regulated by a control loop to maintain a constant desired flow rate. 

The default outlet of the acoustophoresis chip was discarded in a pressurised container, while the sorted outlet was connected to the RPS chip via  a 4.5 mm long PTFE tubing (0.3 mm inner diameter). The flow rate towards the RPS chip was regulated to 3 $\mu$L/min.

\subsection{Acoustic power modulation}

During operation, the SAW transducers were powered with a sinusoidal voltage of amplitude 500 $mV_{pp}$ generated by an arbitrary waveform generator (AWG) gated by an Arduino module and then amplified by a 30 dB PARF310004 power amplifier (ETSA) (see \reffig{fig: general_schematic}).  According to a calibration procedure using acoustic streaming in micro-droplets \cite{riaud2017influence} described in the supplementary information, the standing SAW displacement at the center of the channel was approximately 0.52 nm$_{RMS}$. Literature data indicate that the leaky SAW decays by approximately 0.4 dB/wavelength, that is 2 dB over the channel width \cite{royer_atten,toru2014realisation,jo2014effects}. 

In order to adjust the average magnitude of the acoustic radiation force, we used the duty cycle of the Arduino Power Modulation (PWM). Provided that the gating frequency (490 Hz) is much slower than the SAW frequency, equation \eqref{eq: Force} remains valid. Meanwhile, as long as the modulation period is much shorter than the time particles take to travel across the sorting section, the particles only experience the average acoustic power. 

Besides time-dependence constraints, the choice of power levels during the acoustophoresis was further guided by two aspects: (i) the power modulation frequency of the Arduino chip is close to the lock-in amplifier, which adds noise to the measured signal from the sensor chip. In order to increase the  signal-to-noise ratio (SNR), we alternated periods of 5 s on and 5 s off. This 5 s duration was chosen much smaller than the characteristic duration of the impulse response (approximately 50 s, see the results section) and much longer than the residence time of the particles in the microchannel. All the RPS measurements were conducted during the 5 s off, and the 5 s on samples were discarded. (ii) PDMS is a strongly attenuating material that absorbs quickly the SAW power which drives significant temperature increase in the vicinity of the SAW\cite{ha2015acoustothermal,ha2015generation}, hence the power must remain low enough not to perturb significantly the experiment, In preliminary experiments, the PDMS showed evidence of thermal damage when the time-averaged acoustic energy density exceeded 22.8 J/m$^3$ (ie when the duty ratio was above 50\%). Therefore, this duty ratio was selected as the upper bound for subsequent experiments. 

\begin{figure}
\includegraphics[width=80mm]{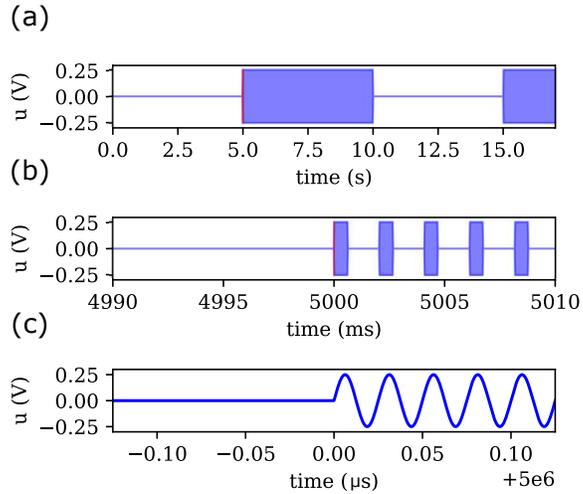}
\caption{Main experiment timescales. \textbf{a}: on and off deviation periods (5 s each) to improve the SNR. \textbf{b}: 490 Hz power modulation controlled by the Arduino chip. The duty cycle controls the average acoustic power experienced by the particles as they travel through the sorting chip. \textbf{c}: 40 MHz sinusoidal wave to generate the SAW. }
\label{fig: timescales}
\end{figure}

The main experiment time scales are illustrated in \reffig{fig: timescales}. In order to measure the compressibility of the particles, the acoustic power was selected randomly within 11 regularly spaced values every 10 s cycle. The exact sequence of power and the detection events were recorded for further analysis.

\subsection{Resistive Pulse Sensing}

The RPS chip was composed of a series of filters to prevent clogging (smallest cross-section 20 $\mu$m) followed by a $L_{RPS}=100$ $\mu$m-long $20\times20$ $\mu$m wide sensing section (see \reffig{fig: schem RPS}). In order to generate a quasi-uniform electric field required by equation \eqref{eq: res. coulter}, we set the distance between the electrodes to 100 $\mu$m. The uniformity of the field was verified using Comsol (data in SI). The electrodes themselves were a symmetric assembly of an a active electrode and two sensing electrodes. Each of the electrodes was 20 $\mu$m wide. The fabrication process is detailed in the supplementary information.

We probed the channel resistance by powering the Wheatstone bridge with a 5 $V_{pp}$ AC-excitation at 50 kHz followed by lock-in demodulation with a gain of 1,000 (SCITEC 441).

\section{Results and discussion}

\begin{figure}
\includegraphics[width=80mm]{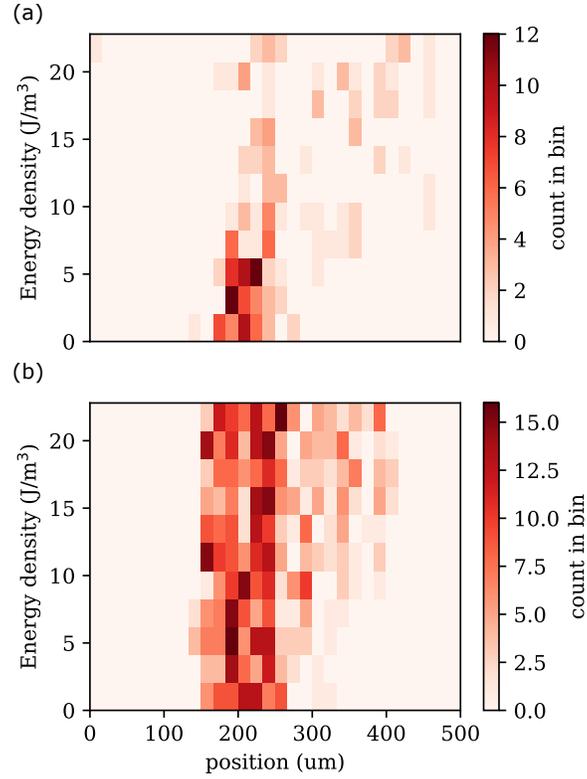}
\caption{Video analysis of particles deflection for a range of power levels. The $x$ axis represents the microspheres \textbf{(a)} or cells  \textbf{(b)} position just before exiting the sorting channel, while the $y$ axis indicates the relative power level $\mathcal{E}$. For each bin delimiting a power level ($\pm4.5\%$) and position (($\pm8.3$ $\mu$m), the number of particles or cells leaving the channel over a 1s interval is indicated by the color scale.}
\label{fig: video_analysis}
\end{figure}

In order to clearly compare the current method with earlier approaches based on video analysis, we first recorded the deflection of particles at the outlet of the sorting section over a range of power levels. Video analysis indicates that sorting occurs if the particle position in the channel exceeds $382 \pm 40$ $\mu$m. Hence, according to \reffig{fig: video_analysis} the polystyrene microspheres are sorted for power levels exceeding $\mathcal{E} = 11.4$ J/m$^3$, whereas the cells have no clear threshold even though some deflection occurs as early as $\mathcal{E} = 9.1$ J/m$^3$.

\begin{figure}
\includegraphics[width=60mm]{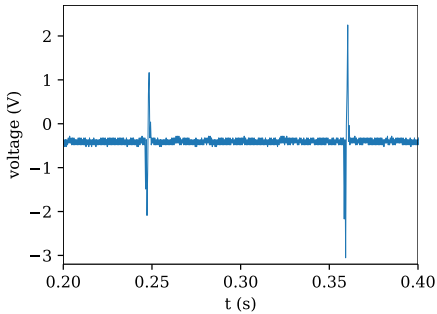}
\caption{Waveform trace recorded from the RPS chip after amplification. The smaller peak on the left is a single PS particle while the peak on the right cannot be identified without knowledge of the acoustic power level.}
\label{fig: sig_sample}
\end{figure}

Next, we compare this video analysis to the RPS chip output. A typical signal trace (after amplification) is shown in \reffig{fig: sig_sample}. Each half peak lasts approximately 2.3 ms. In the experiment, we used the 7.32 $\mu$m diameter microspheres (V = 205 $\mu m^3$) as a calibration standard to establish the voltage-volume relation coefficient, and obtained 170 $\mu m^3$/V. The experimental standard deviation of the particle size distribution is inferior to the data from manufacturer (see SI), which supports that our design does not introduce additional bias due to the vertical position of the particles in the channel \cite{grenvall2014two}. 

\begin{figure}
\includegraphics[width=80mm]{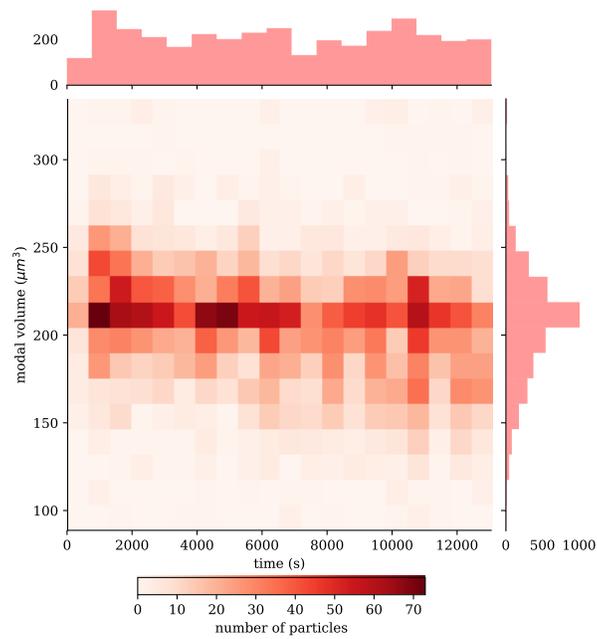}
\caption{Arrival rate of particles with a volume ranging from 150  $\mu$m$^3$ up to 350 $\mu$m$^3$. The heatmap at the center indicates the number of particles counted in the time interval $t\pm 324$ s with a volume $V\pm 37$ $\mu m^3$. The bar graph at the top represents the arrival time distribution of the entire population of particles, while the bar graph on the right represents the volume distribution of the entire population of particles regardless of their arrival time.}
\label{fig: PS_histo}
\end{figure}

Over the course of 4 hours, we recorded the arrival rate of particles. Except for very few outliers, most particles volume ranged from 150  $\mu$m$^3$ up to 1,600 $\mu$m$^3$. \reffig{fig: PS_histo} is a composed histogram showing the number of particles with a volume smaller than 350 $\mu$m$^3$ arriving over time intervals (75 s) much larger than the power fluctuation period (10 s). This attenuates the stochastic nature of the power sequence and demonstrates that the particles arrive in the detector at a constant rate (top histogram). This population of particles is further subdivided into smaller groups of identical volumes which yields the center two-dimensional histogram. This graphic shows a downward trend that indicates a slight decay in the average detected particle volume over time (larger particles sediment faster). Finally, the histogram on the right indicates the total number of particles detected over 4 hours and is representative of the size distribution of the PS microspheres.

\begin{figure}
\includegraphics[width=80mm]{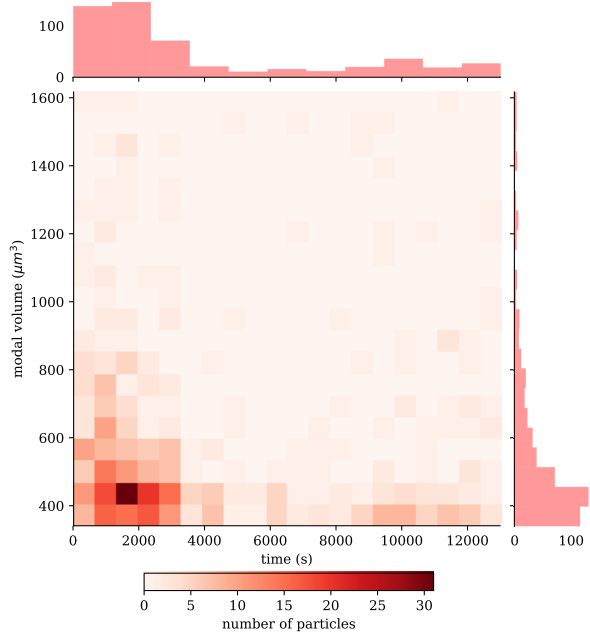}
\caption{Arrival rate of particles with a volume ranging from 350  $\mu$m$^3$ up to 1,600 $\mu$m$^3$. The heatmap at the center indicates the number of particles counted in the time interval $t\pm 324$ s with a volume $V\pm 32$ $\mu m^3$. The histogram at the top represents the arrival time distribution of the entire population of particles, while the histogram on the right represents the volume distribution of the entire population of particles regardless of their arrival time.}
\label{fig: cells_histo}
\end{figure}

Unlike these small PS beads, the arrival rate of larger objects decays much faster  as shown in \reffig{fig: cells_histo}. We believe this decay is due to the sedimentation of the particles despite a continuous stirring of the liquid reservoir\cite{baret2009remote}. According to the literature, Jurkat cells diameter (volume) ranges between 10.5 and 12.5 $\mu m$ (600 up to 1000 $\mu m^3$). Hence, the objects with a volume below 600 $\mu m^3$ are suspected to be PS microspheres doublet. This hypothesis will be verified thereafter using compressibility data. 

\subsection{Calculation of the deviation thresholds}

In the previous section, we have exposed a stream of microspheres and cells to increasing levels acoustic power until a deviation threshold $\mathcal{E}_{min}$ was reached, at which point the microspheres and then the cells started to be deflected towards the analysis chip. According to Eq. \ref{eq: Pelton}, the deviation only depends on the number $M$ (Eq. \eqref{eq: M}), therefore the onset of sorting yields exactly $|M|=|\sin\theta_C|$. 
However, knowing $M$ is not enough to immediately deduce the particle acoustic properties. Besides acoustic contrast, $M$ also depends on external factors such as the flow velocity, fluid viscosity and acoustic energy density, but also on the particle radius. The latter is directly evaluated with the RPS sensor (assuming a spherical geometry for the particles).  Hence, the acoustic contrast is obtained from:
\begin{equation}
\Phi_\mathtt{exp} = \frac{9\eta v_F|sin\theta_C|}{2a^2k_X\mathcal{E}_{min}} \label{eq: phi_exp}
\end{equation}
Once the acoustic contrast is known, recovering $f_1$ and $f_2$ (knowing the particle density) is straightforward. 

In the following, we first use the linear method developed above to recover the deviation threshold for the polystyrene microspheres, then apply the statistical methods to analyze less abundant particles. Eventually, the acoustic contrast of each type of particle is recovered and the particle compressibility is computed.

\subsubsection{Linear method}

\begin{figure}
\includegraphics[width=80mm]{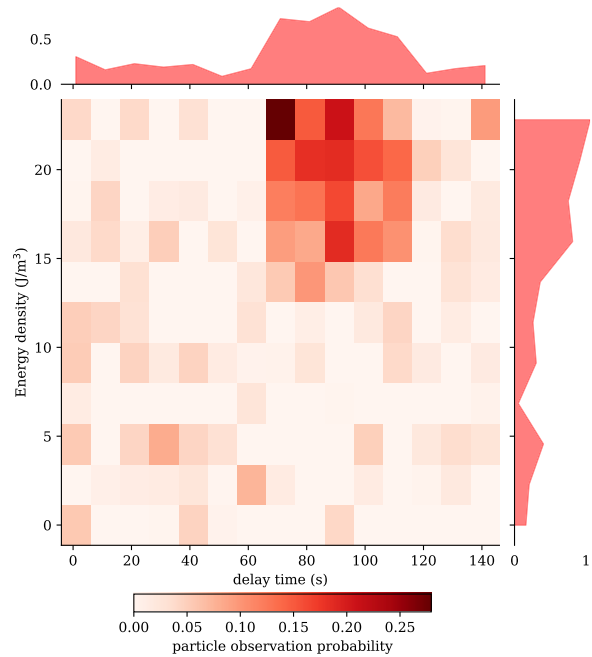}
\caption{PS particles (7.32 $\mu$m diameter) sorting impulse response estimated from the linear regression solution $G$. The heatmap at the center shows the impulse response as a function of the power level and the delay after application of the pulse. The histogram on the top represents the cumulated impulse response across all power levels while the histogram on the right indicates the likelihood of deviation at a given power level.}
\label{fig: linear_sol}
\end{figure}

The system $\|\mathcal{D}G-B\|^2$ (Eqs. (\ref{eq: linear system D}-\ref{eq: linear system B})) is minimized with the nonzero least square solver of Octave. The reshaped version of $G$ is presented in \reffig{fig: linear_sol}.

According to \reffig{fig: linear_sol}, the PS microspheres take between 70 s and 120 s to flow along the 4.5 mm tubing between the sorting and the sensing sections. Almost no particles are sorted unless the relative acoustic energy $\mathcal{E}$ exceeds 13.7 J/m$^3$ which is comparable to the results from the video analysis (11.4 J/m$^3$).

\subsubsection{Statistical method}

When the number of particles is too low, the linear model yields inaccurate results, so the statistical model should be used instead. We recall that this model studies the difference of average acoustic energy density in the moments that precede the detection of a particle (as compared to when no particle was detected). The analytical formula from Eq. \ref{eq: stat} is first validated against numerical simulations, then we assess the robustness of the method against statistical noise, and then use this method to analyze experimental results.

\begin{figure}
\includegraphics[width=80mm]{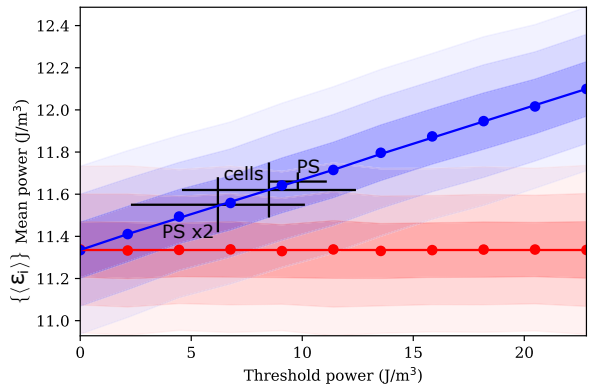}
\caption{Statistical inference of the particle deviation threshold. We simulated 1,000 independent experiments in which $N=200$ particles are observed, and for each particle the power is averaged over $n_t=15$ discrete time steps. Of these 15 time steps, only 1 is expected to determinate the sorting result of the particle. The solid blue (red) line represent the analytical result given by equation \eqref{eq: stat}, and the dots show the population and time-averaged power levels for the particles (control) averaged over 1,000 simulations. Each degree of shaded blue (red) areas indicate a standard deviation between simulated experiments. The crosses represent the experimental results from a single experiment involving 200 cells or particles doublet and more than 3,000 PS particles. The width of the cross is the standard deviation obtained from the simulations using similar population sizes.}
\label{fig: mean_shift_analysis}
\end{figure}

In order to validate the statistical method, we simulated the sorting and detection of particles with a hard threshold below which no particles are sorted and above which all the particles are sorted. Since the PS microspheres were detected within 150 s (15 random power levels), we also used $n_t=15$ in these simulations. The results of 200,000 simulations of sorting events and 200,000 negative controls are depicted in \reffig{fig: mean_shift_analysis} by the blue and red dots respectively. The linear trend of these dots compares well to the analytical formula from Eq. \ref{eq: stat} (solid lines). 

Due to the statistical nature of the model, we then wanted to evaluate its reliability: the 200,000 simulations were grouped into 1,000 sets of 200 particles, yielding 1,000 possible outcomes. The standard deviation between these outcomes is also presented in \reffig{fig: mean_shift_analysis} in a series of shades. Each shade represents a standard deviation between the 1,000 outcomes. This process was repeated for each discrete level of acoustic power used in our experiments, which yields the bands shown in \reffig{fig: mean_shift_analysis}. According to the simulations, differences in sorting threshold above 5 J/m$^3$ should exceed a standard deviation.

The experimental results of the PS microspheres, cells and PS microsphere doublets are also reported on \reffig{fig: mean_shift_analysis} with the associated error bars based on the standard deviation from the simulations (adjusted based on the number of observed particles in each case). The three types particles have a deviation threshold that differs by less than a standard deviation. Nonetheless the size of the particles has not yet been accounted for. These thresholds can also be compared to the video analysis and the physical model used earlier to analyze the deviation of PS microspheres. This statistical approach yields a deviation threshold of 9.8 J/m$^3$, which is reasonably close to the video analysis (11.4 J/m$^3$) and lower than the physical model (13.4 J/m$^3$). Yet, all three methods agree within 20\%.

\subsection{Calculation of scattering coefficients}

\begin{figure}
	\includegraphics[width=80mm]{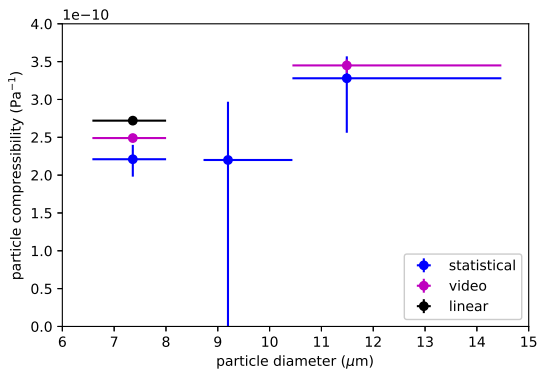}
	\caption{Experimental particle size and acoustic contrast. The analysis method is indicated by the symbol color, with ``stats" standing for ``statistical".}
	\label{fig: final_results}
\end{figure}

Once the particle size and deviation threshold are known, we use Eq. \eqref{eq: phi_exp} to recover the particles acoustic contrast. $f_1$ and $f_2$ are then immediately obtained which allows computing, the particles compressibility. The results are presented in Table \ref{tbl: results} and \reffig{fig: final_results}.

The polystyrene microspheres compressibility (obtained from the statistical method) is very close to the tabulated value\cite{hartono2011chip} of $2.2\times10^{-10}$ Pa$^{-1}$. The microspheres doublets compressibility is estimated to be $2.20\times10^{-10}$ Pa$^{-1}$, which is almost the same as for the PS microspheres. We also note that the $f_2$ coefficient (that depends on the density ratio) is generally much smaller than the compressibility-related $f_1$ coefficient. Hence, the exact value of the particle density is not critical for the results accuracy. 
Even though the cells and the doublets had a similar deviation threshold, the larger size of the cells yields a very different compressibility ($3.28\times10^{-10}$ Pa$^{-1}$). Furthermore, despite the small number of cells detected during the experiment, the estimated compressibility of Jurkat cells is consistent with earlier studies and intermediate between red blood cells ($\kappa_p = 3.18\times10^{-10}$ Pa$^{-1})$ and MCF-12A ($\kappa_p = 3.54\times10^{-10}$ Pa$^{-1})$\cite{hartono2011chip}. 

\begin{table}
	\begin{tabular}{lccccccc}
		\hline
		                           & $\rho_p$ & $\mathcal{E}^{(a)}_{min}$&  $V_p$  & $\Phi_\mathtt{exp}$ & $f_1$ & $f_2$ & $\kappa_p$   \\
		particle                   &(kg/m$^3$)& (J/m$^3$)           & ($\mu$m$^3$) &        &       &       & ($\times 10 ^{10}$, Pa$^{-1}$)   \\
		\hline
		PS (video)                 & 1062     &  $11.4$             & 210          & 0.398  & 0.441 & 0.0397 & 2.49\\
		PS (linear)                & 1062     &  $13.4$             & 210          & 0.339  & 0.390 & 0.0397 & 2.72\\
		PS (statistical)           & 1062     &  $9.8\pm1.1$        & 210          & 0.462  & 0.505 & 0.0397 & 2.21\\
		\hline
		cells (video)              & 1075     &  $10.2\pm0.9$       & 800          & 0.175  & 0.226 & 0.0476 & 3.45\\
		cells (statistical)        & 1075     &  $8.5\pm3.6$        & 800          & 0.213  & 0.264 & 0.0476 & 3.28\\
		\hline
		PS $\times$2 (statistical) & 1062     &  $6.2\pm3.6$        & 400          & 0.472  & 0.515 & 0.0397 & 2.20\\
		\hline
	\end{tabular}
\caption{Analysis of cells and particles deviation. (a) acoustic energy density estimated from the 0.52 nm$_{RMS}$ displacement.}
\label{tbl: results}
\end{table}

Our final results are synthesized in \reffig{fig: final_results}. The three methods (video, linear and statistical) yield slightly different results for the compressibility of PS microspheres. Nonetheless, PS microspheres are clearly distinct from other kinds of particles in terms of size and can be identified with certainty. The cells and PS doublets show a slight overlap in size and compressibility, but the combination of both parameters lifts the ambiguity and indicates more clearly that these two populations do refer to two different types of particles. This result could not have been obtained from any of these two methods alone.

\section{Perspectives}

This work marks a first step towards the integration of RPS sensing and acoustophoresis on a single chip. Since the proposed method does not require high-speed camera or microscopes, a natural continuation would be to integrate the acoustophoresis and RPS in a single chip for point-of-care diagnostic. From a more fundamental point of view, acoustophoresis and RPS are not restricted by the diffraction limit, hence combining these two technologies may allow probing the mechanical properties of nano-objects such as nanoparticles, exosomes and viruses. Nonetheless, the current system still faces several challenges that need to be addressed before the technology reaches its full potential for point-of-care applications and nanoparticle analysis.

In our opinion, the two major limitations of the current device are (i) that it is not yet truly single-cell and (ii) that the density of the particles has to be calibrated in a different experiment. The single-cell limitation stems from the delay between sorting and detection, which is mostly due to the tubing interconnect between the sorting and RPS chips. This tubing generates a Taylor diffusion such that individual detection events cannot be directly linked to the acoustic energy density. This issue should disappear once the sorting and sensing functions are integrated on a single chip (thereby eliminating the tubing and thus the delay and need for probabilistic models). Since both chips are fabricated using the same process on similar substrates, such integration may be within reach. 
Regarding the need to know the particle density, we anticipate two approaches. According to Eq. \eqref{eq: SAW contrast factor}, choosing $\theta_R=45^o$ makes $\Phi$ independent of the particle density. Such Rayleigh angle can be achieved by lowering the SAW velocity, for instance by switching material or using thinner substrates. An alternative approach would be measuring the speed of sedimentation in a configuration similar to Grenvall \etal \cite{grenvall2014two}. In this work, the authors showed that the vertical position of the particles can be sensed by an RPS system with neighboring electrodes, thus a pair of such electrodes can measure the sedimentation speed of the particles and thus their density. %Reversely, depending on the assumptions required for each study, it may be possible to use modified media to ensure that most particles are neutrally buoyant. Even though this could solve some of the sedimentation issues faced in current study, we believe it could complicate the establishment of a measurement standard and decided to keep unmodified media. 
Besides density, compressibility and size, higher-end lock-in amplifier may also allow to record the cell electrical impedance at various frequencies with a similar setup. 

%Both RPS and acoustophoresis can be biocompatible. Some future work could also use the combination of devices to sort the cells based on those mechanical criteria.

\section{Conclusion}
Resistive pulse sensing has long been limited for the measurement of cells and particles mechanical properties.  In this work, we used acoustophoresis to provide the mechanical insight to the RPS. This required several theoretical and technological advances, including studying the deviation of particles in a tilted-angle acoustic field, designing a modular two-chips experiment and accounting for the time delay between particle sorting and detection when analyzing the data. Our results were scrutinized by three different methods, which approximately agreed on the particle compressibility. Furthermore, in contrast to constriction-based methods, mechanical phenotyping can be performed over a much broader range of particle size and elasticity, including cells and solid particles. With further integration, this strategy could yield point-of-care mechanical phenotyping devices and allow analyzing nanoparticles, exsosomes and viruses.


\begin{acknowledgement}
The authors gratefully acknowledge Gabriele Pitingolo, Shufang Renault and Leonard Jagot Lagoussiere for their useful discussion, Aloysa Guerra and Catherine Dode for the generous gift of Jurkat cells, Philippe Nizard for his help with cell manipulation and Michael Baudoin for his fruitful discussions.This research was supported by the Fondation pour la Recherche Medicale (SPF20160936257)
\end{acknowledgement}

\begin{suppinfo}

Fabrication of the microchannels, fabrication of the interdigitated transducers, calibration of the SAW power and
evaluation of the electric field homogeneity in the sensing channel.

\end{suppinfo}

\bibliography{Coulter_counter}

%\pagebreak
%\section{Graphical TOC Entry}

\end{document}